\documentclass[12pt,letterpaper]{article}
\usepackage[T1]{fontenc}
\usepackage{geometry}
\usepackage{setspace}

\usepackage[style = chem-acs]{biblatex}
\usepackage{graphicx}
\usepackage{float}
\newfloat{scheme}{htbp}{los}
\floatname{scheme}{Scheme}
\floatname{chart}{Chart}
\newfloat{graph}{htbp}{loh}

\usepackage{chemformula} 

\renewcommand{\r}{\vec{r}}

\newcommand{\ket}[1]{\left|#1\right>}
\newcommand{\bra}[1]{\left<#1\right|}

\newcommand{\bak}[3]{\left<#1|#2|#3\right>}

\usepackage{authblk}
\author[1]{Benjamin G. Janesko}
\affil[1]{Department of Chemistry \& Biochemistry, Texas Christain University, Fort Worth, TX 76129, USA}

\title{Restoring the uniform density limit in Perdew-Zunger self-interaction correction}
\date{*Email: b.janesko@tcu.edu}

\begin{document}
\onehalfspace

\maketitle

\begin{abstract}
The Perdew-Zunger self-interaction correction (PZ-SIC) makes approximate
density functional theory (DFT) exact for all one-electron densities, but
sacrifices exactness for uniform densities. I show that an alternative to the
orbital density {\em{ansatz}} employed in PZ-SIC restores the uniform density
limit.  The new {\em{ansatz}} can be expressed as a local hybrid
functional including an orbital-dependent local mixing functional. This
approach  also eliminates the need to evaluate approximate density functionals
on lobed one-electron densities extracted from smooth many-electron densities,
thereby reducing orbital dependence and lobed density error. I demonstrate the
alternative {\em{ansatz}} in a broadly accurate nonempirical locally scaled
self-interaction-corrected generalized gradient gradient approximation 
\end{abstract}

\section*{Keywords}

Density functional theory; self-interaction correction; local hybrid


\section{Introduction}

Kohn-Sham density functional theory (DFT) remains the most widely used solution
to the many-electron problem.\cite{Janesko2021} DFT models the ground-state
energy and density of a real N-electron system in terms of a reference system
of N noninteracting electrons, corrected by a mean-field Hartree interaction
and a formally exact exchange-correlation (XC) density functional which combine
to recover the electron-electron interaction energy.\cite{Perdew2009} Progress
in DFT is driven by progress in approximate XC functionals.  The local
spin-density approximation (LSDA) models the XC energy density at point $\r$ as
a function of the spin-up and spin-down electron densities $n_\sigma(\r)$,
where $n_\sigma(\r)d^3\r$ is the probability of finding a
$\sigma=\uparrow,\downarrow$-spin electron in infinitesimal volume $d^3\r$
about $\r$. Semilocal (meta) generalized gradient approximations ((m)GGAs)
incorporate local density derivatives and noninteracting kinetic energy
densities \& density Laplacians. 
Approximate functionals can be constructed to obey exact
conditions. 
The present work focuses on three canonical nonempirical approximate
functionals: the LSDA, the Perdew-Burke-Ernzerhof GGA,\cite{Perdew1996} and the
strongly constrained and appropriately normed mGGAs SCAN and r$^2$SCAN. The
latter satisfy 17 exact conditions on the XC
energy.\cite{Sun2015,Bartok2019,MejiaRodriguez2019,Furness2020} 

\subsection{Self-interaction error and its correction} 

All semilocal functionals violate an important exact condition, incorrectly
predicting nonzero electron-electron interaction energies for general
one-electron densities.  This self-interaction error (SIE) manifests as
over-delocalization of charge and spin and is arguably the greatest outstanding
challenge in DFT.\cite{Ruzsinszky2006,Toher2005,Hofmann2012,Kim2013,Burgess2024,Bryenton2022}
The Perdew-Zunger self-interaction correction (PZ-SIC) restores this exact
condition by removing self-interaction on an orbital-by-orbital
basis.\cite{Perdew1981}  Size-extensive PZ-SIC requires using spatially localized
transformed occupied orbitals.\cite{Pederson1984} Limiting this transformation
to Fermi-L\"owdin orbitals (FLOSIC) has made the PZ-SIC widely
available,\cite{Pederson2014,Yang2017,Jackson2019} with recent applications to
water,\cite{Sharkas2020,Dasgupta2021} water-ion \& ammonia
clusters,\cite{Wagle2021,Ufondu2023} polarizabilities,\cite{Akter2021} and
spin-crossover complexes.\cite{Ruan2023}
Self-interaction and the PZ-SIC have emerged as foundational concepts in DFT,
motivating work on the derivative discontinuity and piecewise
linearity,\cite{Perdew1982,Cohen2008a,Bajaj2017,Bajaj2022,Baerends2019,Baerends2022}
Koopmans-compliant DFT and the role of self-interaction in Kohn-Sham orbital
energies,\cite{Dabo2010} and more. 
 
Local hybrid functionals provide an alternative treatment of self-interaction
error.\cite{Jaramillo2003,Maier2018} Hartree-Fock theory is self-interaction
free, as are the closely related exact-exchange Kohn-Sham and generalized
Kohn-Sham methods.\cite{Sala2001,Perdew2017} Admixture of exact exchange to an
approximate XC functional can reduce the SIE.\cite{Becke1993} Local hybrids introduce a
position-dependent admixture of the energy density corresponding to HF exact
exchange.\cite{Jaramillo2003} An appropriate choice of the local mixing
function ensures exactness for any one-electron density.\cite{Johnson2014}
The connectiosn between local hybrids and locally scaled PZ-SIC have drawn
recent attention.\cite{Shahi2026} This work makes the connection explicit,
expressing the alternative SIC {\em{ansatz}} discussed here as a local hybrid
with an orbital-dependent local mixing function. 

\subsection{The uniform limit and lobed density error}

Unfortunately, the PZ-SIC fixes one exact condition by violating another.
Uniform densities, and the related large-nuclear-charge (large-Z) limit of
neutral atoms, provide an important ``appropriate norm'' for XC functional
construction.\cite{Sun2015} Practically all nonempirical semilocal XC approximations
(as well as many empirical ones) are exact for uniform densities. The PZ-SIC
reintroduces errors for these systems, around 3-5\% in the XC energy per
electron of uniform systems.\cite{Santra2019a} 
Violation of this appropriate norm is connected to a ``lobed density error''
intrinsic to Perdew \& Zunger's {\em{ansatz}},  arising when semilocal density
functionals are evaluated on lobed orbital densities carved out of smooth
many-electron densities.\cite{Kluepfel2011,Hofmann2012a,Shahi2019} 
Choosing the localized orbitals to be complex reduces but does not eliminate
these errors.\cite{Kluepfel2011}
 
Violation of the uniform and high-density limits has been invoked to explain
the ``paradox'' of PZ-SIC: correcting SIE can often worsen a semilocal
functional's good description of equilibrium properties such as atomization
energies.\cite{Perdew2015,Zope2019,Akter2021} Such zero-sum tradeoffs between
over-delocalization and underbinding are not restricted to PZ-SIC, and appear
to to arise in any approximate functional when self-interaction mimics electron
correlations in bonded systems.\cite{Janesko2017,Janesko2021} 
Previous efforts to restore the uniform limit involved scaling down the SIC to
zero for uniform systems.\cite{Vydrov2006,Zope2019,Bhattarai2020,Shahi2026} To
date, these efforts have not provided a definitive solution to the paradox of
SIC. 

In this work, I show that an alternative to the orbital density {\em{ansatz}}
employed in PZ-SIC restores the uniform and large-Z limits. This  alternative
local-hybrid-like {\em{ansatz}} can be explicitly written as a local hybrid
functional with an orbital-dependent local mixing function. The {\em{ansatz}},
originally derived for a wavefunction-in-DFT generalization of
PZ-SIC,\cite{Janesko2022d,Janesko2022b,Janesko2025a} also reduces the orbital
dependence and lobed density error by removing the need to evaluate semilocal
functionals on lobed orbital densities. 

\section{Theory} 

Consider a semilocal DFT treatment of
$N=\sum_{\sigma=\uparrow,\downarrow}N_\sigma$ electrons in external potential
$\hat{v}_{ext}$. Suppose that the noninteracting Kohn-Sham reference system is
described by a single-Slater-determinant wavefunction $\ket{\Phi}$, occupied
Kohn-Sham orbitals $\{\psi_{i\sigma}(\r)\}$, spatially localized orthonormal
transformed orbitals $\{\phi_{i\sigma}(\r)\}$, and spin densities
$n_{\sigma}(\r)=\sum_{i=1}^{N\sigma}\left|\phi_{i\sigma}(\r)\right|^2$. The
Kohn-Sham energy is 
\begin{eqnarray}
E_{KS} &=& \bak{\Phi}{\hat{h}_0}{\Phi}+U[n]+ \int d^3\r\ e_{XC}^{app}[n_\uparrow(\r),n_\downarrow(\r)]
\end{eqnarray}
Hartree atomic units are used throughout. Operator
$\hat{h}_0=\hat{T}+\hat{v}_{ext}$ where $\hat{T}$ is the electron kinetic
energy operator. The Hartree energy is $U[n]=(1/2)\int d^3\r\int
d^3\r\,'n(\r)n(\r\,')|\r-\r\,'|^{-1}$. The semilocal approximate XC energy
should (but does not) cancel the Hartree energy for any one-electron density.

The PZ-SIC energy is 
\begin{eqnarray}
\label{eq:1}
E &=& E_{KS} + \sum_{i\sigma} \int d^3\r\ f_\sigma\left(e_{Xi\sigma}(\r)-e_{XC}^{app}[n_{i\sigma},0]\right)
\end{eqnarray}
Here $n_{i\sigma}=|\phi_{i\sigma}(\r)|^2$ is the orbital density. The exact
self-interaction of transformed spinorbital $\phi_{i\sigma}$ can be obtained
from its exact exchange energy density
\begin{eqnarray}
\label{eq:exi}
e_{Xi\sigma}(\r)&=&-\frac{1}{2}\int
d^3\r'\frac{|\phi_{i\sigma}(\r)|^2|\phi_{i\sigma}(\r\,')|^2}{|\r-\r\,'|}
\end{eqnarray}
Integrating this quantity over all space yields minus the
self-Coulomb energy $U[n_{i\sigma}]$.  Weight $f_\sigma$ is 1 in PZ-SIC and
position dependent in locally scaled
SIC.\cite{Zope2019,Bhattarai2020,Shahi2026} The last term in eq \ref{eq:1} is
the orbital density {\em{ansatz}} introduced by Perdew and Zunger. 
In the Perdew-Zunger {\em{ansatz}}, one must evaluate the approximate XC energy
functional on the one-electron orbital densities. This choice appears to be
respondible for the violation of uniform and large-Z limits, and is definitely
responsible for the lobed density errors discussed above.\cite{Shahi2019} 

\subsection{Local-hybrid-like self-interaction correction LHSIC} 

This work treats an alternative {\em{ansatz}} developed for a
wavefunction-in-DFT generalization of
PZ-SIC.\cite{Janesko2022d,Janesko2022b,Janesko2025a}
\begin{eqnarray}
\label{eq:2}
E_{LHSIC} &=& E_{KS} + \sum_{i\sigma} \int d^3\r\ f_\sigma \Big(e_{Xi\sigma}(\r) \\ 
&&-\frac{e_{Xi\sigma}(\r)}{e_{X\sigma}(\r)}e_{XC}^{app}[n_\uparrow(\r),n_\downarrow(\r)]\Big) \nonumber 
\end{eqnarray}
This local-hybrid-like self-interaction correction (LHSIC) incorporates the
exact exchange energy density employed in local hybrids and other
hyper-generalized-gradient (hyper-GGA) functionals\cite{Jaramillo2003,Kaupp2024,Perdew2008,Haunschild2012}
\begin{eqnarray}
e_{X\sigma}(\r)&=&-\frac{1}{2}\sum_{ij}\int
d^3\r\,' \frac{\psi^*_{i\sigma}(\r)\psi^*_{j\sigma}(\r')\psi_{j\sigma}(\r)\psi_{i\sigma}(\r')}{|\r\,'-\r|}
\end{eqnarray}
The LHSIC also incorporates the exact exchange energy densities of each orbital $i$ introduced above. 
Eq \ref{eq:2} scales the full XC energy density by the ratio of orbital exact
exchange $e_{Xi\sigma}(\r)$ and full exact exchange $e_{X\sigma}(\r)$. 
Eq \ref{eq:2} is exact by construction for any one-electron density and
recovers the large-Z and high-density uniform limits where
$e_{XC}^{app}/e_X\to 1$. Replacing
$e_{XC}^{app}$ in eq \ref{eq:2} with the exchange-only piece ({\em{cf.}} eq
\ref{eq:llh} below) restores the exact limit for all uniform densities and all
one-electron densities when paired with a self-interaction-free meta-GGA
correlation functional.\cite{Sun2015}

\subsection{Local hybrid formulation of LHSIC } 

The LHSIC {\em{ansatz}} in eq \ref{eq:2} can be written explicitly as a local
hybrid functional possessing a orbital-dependent mixing function. The general
form of a local hybrid functional is 
\begin{eqnarray}
\label{eq:lh} 
E_{LH} &=& E_{KS} + \sum_\sigma \int d^3\r a_\sigma(\r)\left(e_{X\sigma}(\r) -e_{X\sigma}^{app}[n_\sigma(\r)] + G(\r)\right)
\end{eqnarray}
The local mixing function (LMF) $a_\sigma(\r)$ controls the admixture of exact
exchange. 
Common LMFs $a_\uparrow(\r)=a_\downarrow(\r)$ can provide
beyond-zero-sum performance for diverse properties.\cite{Arbuznikov2009a} 
Choosing the LMF to be one in one-electron regions restors the exact limit for
all one-electron densities  when paired with a self-interaction-free meta-GGA
correlation functional.\cite{Sun2015} One such choice is the
t-LMF\cite{Jaramillo2003,Bahmann2007,Arbuznikov2007}
\begin{eqnarray}
\label{eq:twt}
a_\sigma^{tLMF}(\r)&=&\frac{\tau_{W\sigma}(\r)}{\tau_\sigma(\r)}
\end{eqnarray}
Here 
$\tau_\sigma(\r)=(1/2)\sum_i|\nabla\psi_{i\sigma}(\r)|^2$ and $\tau_W$ denote
the exact (positive-semidefinite) and von Weiszacker kinetic energy densities.
The energy gauge correction in eq \ref{eq:lh} obeys $\int d^3\r G(\r)=0$ and is
intended to correct discrepancies between the gaugues of exact vs. semilocal
approximate exchange.\cite{Perdew2008,Theilacker2016,Arbuznikov2024,Maier2016a}
Recently, an x-LMF incorporating a dependence on the exact exchange energy
density itself was shown to reduce the need for energy gauge correction.\cite{Arbuznikov2024}
\begin{eqnarray}
\label{eq:xlmf} 
a^{xLMF}(\r) &=& {\textrm{erf}}\left[\eta {\textrm{exp}}\left(\vartheta\left(w(\r)-1\right)\right)\right] \\ 
w(\r) &=& \frac{e_{X}(\r)}{e_X^{app}(\r)}
\end{eqnarray}
Here for example $e_X(\r)=\sum_{\sigma} e_{X\sigma}(\r)$ denote spin-summed
exchange energies and parameters $\eta$ and $\vartheta$ are fit to molecular
thermochemistry and kinetics. 
Other details of local hybrids have been recently reviewed.\cite{Kaupp2024}

\begin{figure}
\includegraphics[width=\textwidth]{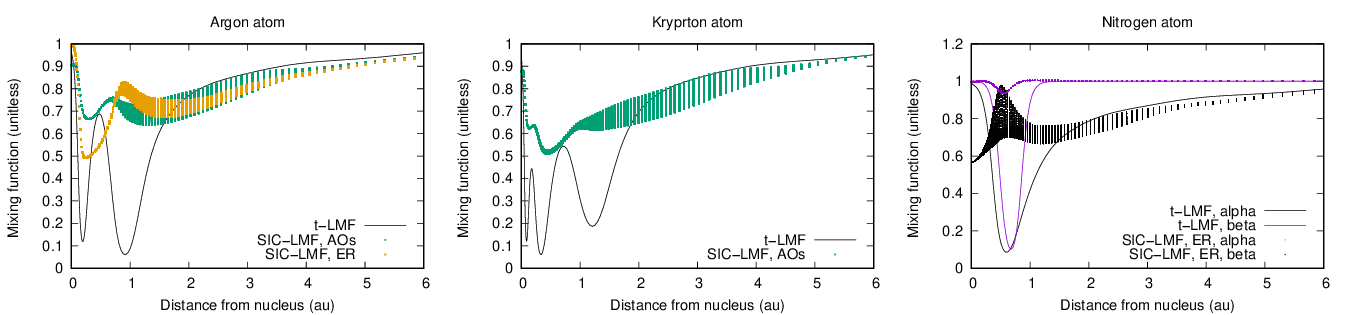}
\caption{\label{fig:lmf} Local mixing functions in argon, krypton, and nitrogen
atoms, plotted versus distance from the nucleus. The orbital-dependent SIC-LMF
breaks spherical symmetry, leading to more than one value at a given distance
from the nucleus.}
\end{figure} 

The key for our purposes is that the local-hybrid-like self-interaction
correction eq \ref{eq:2} can be written as eq \ref{eq:lh} with an
orbital-dependent self-interaction-correcting local mixing function (SIC-LMF):
\begin{eqnarray}
\label{eq:siclmf} 
a_\sigma^{SICLMF}(\r) &=& \frac{\sum_i e_{Xi\sigma}(\r)}{e_{X\sigma}(\r)}
\end{eqnarray}
Here the orbital self-exchange energies are from eq \ref{eq:exi}. The
corresponding common LMF\cite{Arbuznikov2009a} is $a^{SICLMF}(\r)=\left(\sum_{i\sigma}
e_{Xi\sigma}(\r)\right)e_{X}^{-1}(\r)$. 
The SIC-LMF  depends explicitly on the exact exchange energy
density, making it similar to the local mixing functions employed in the
Perdew-Staroverov-Tao-Scuseria hyper-GGA\cite{Perdew2008} and the 
x-LMF.\cite{Arbuznikov2024}
Figure \ref{fig:lmf} compares the SIC-LMF in eq \ref{eq:siclmf} to the
original\cite{Jaramillo2003} t-LMF $\tau_W/\tau$. Computational details are in section
\ref{sec:methods} below. The SIC-LMF is overall similar to the t-LMF in
valence regions beyond about 2 au from the nucleus. The two LMF show
differences in core and core-valence regions. The SIC-LMF breaks spherical
symmetry in spherical atoms possessing a fully occupied $p$ or $d$ shell. This
is clear in spin-quartet nitrogen atom, where the majority-spin SIC-LMF breaks
spherical symmetry while the minority-spin value does not.  The plots for argon
and krypton atoms show the SIC-LMF evaluated from atomic orbitals as well as
Edmiston-Reudenberg (ER) localized orbitals. SIC-LMFs computed from the two
sets of orbitals are rather similar in the valence region and differ in the
core-valence regions, consistent with the relatively modest orbital
localization dependence discussed below. 

\subsection{Rationalizing the local hybrid {\em{ansatz}} from projected-interacting DFT} 
\label{proj}

The local hybrid {\em{ansatz}} in eq \ref{eq:2} can be rationalized by the
projected-interacting derivation of self-interaction
correction.\cite{Janesko2022b,Janesko2025a,Janesko2025} I have proposed
projected-interacting DFT as a generalization of Kohn-Sham DFT, closely related
to range-separated wavefunction-in-DFT methods.\cite{Savin1995,Leininger1997,Savin2020,Toulouse2004} 
 
Range-separated wavefunction-in-DFT generalizes Kohn-Sham DFT by
reintroducing {\em{part of}} the electron-electron interaction operator into
the noninteracting Kohn-Sham reference system. For example, in long-range
wavefunction-in-DFT, one computes a correlated {\em{ab initio}} approximation
for a long-range-interacting reference system's full CI wavefunction, corrected by an
approximation for a formally exact short-range Hartree-exchange-correlation density
functional.\cite{Toulouse2004} Given the exact (full CI)  reference system
wavefunction and the exact short-range functional, the reference system's
ground-state density and energy will equal those of the real system.
 
Projected-interacting wavefunction-in-DFT introduces {{multiple}} different
reference systems, each experiencing a different projected electron-electron
interaction operator. Each projected-interacting reference system
yields the exact ground-state energy and density when corrected by the corresponding exact 
projected Hartree-exchange-correlation density
functional.\cite{Janesko2022b,Janesko2026}

We can recover SIC in the projected-interacting framework by defining a
``self-interacting'' reference system whose projected electron-electron
interaction $\hat{V}^P_{ee}$ is projected onto the SIC spinorbitals introduced
above\cite{Janesko2022b}
\begin{eqnarray}
\label{eq:VP}
\hat{V}^P_{ee} &=& \sum_{i\sigma}\ket{\phi_{i\sigma}\phi_{i\sigma}}\bak{\phi_{i\sigma}\phi_{i\sigma}}{\hat{V}_{ee}}{\phi_{i\sigma}\phi_{i\sigma}}\bra{\phi_{i\sigma}\phi_{i\sigma}}
\end{eqnarray}
The electron-electron interaction energy of this ``self-interacting'' reference
system is zero, $\bak{\Phi}{\hat{V}^P_{ee}}{\Phi}=0$. This energy may be
expressed as a sum of projected Hartree and projected XC energies (which sum to
zero). The projected Hartree energy is the difference between the total Hartree
energy and the self-Hartree energy as defined in PZ-SIC. Critically, the
projected XC energy depends on the {\em{nondiagonal}} projected
electron-electron interaction operator
$\bak{\r_1,\r_2}{\hat{V}^P_{ee}}{\r_1\,' ,\r_2\,'}$. 
("Nondiagonal" means
that $\bak{\r_1,\r_2}{\hat{V}^P_{ee}}{\r_1\,',\r_2\,'}$ does not scale as
$\delta(\r_1-\r_1\,')\delta(\r_2-\r_2\,')$.)
Expectation values of this operator must be evaluated using the
nondiagonal XC hole $h_{XC}[n](\r_1,\r_2;\r_1\,',\r_2')$ 
To illustrate, if we choose to express the exchange holes about reference point
$\r_1$, the diagonal part of the $\sigma$-spin exact exchange hole is
$h_{X_\sigma}(\r_1,\r_2)=h_{X\sigma}(\r_1,\r_1;\r_2,\r_2) =
-(1/2)n_\sigma^{-1}(\r_1)|\gamma_\sigma(\r_1,\r_2)|^2$ and the full nondiagonal
exact exchange hole is 
\begin{eqnarray}
\label{eq:nondiag}
h_{X\sigma}(\r_1,\r_1';\r_2,\r_2') &=& -\frac{1}{2}n_\sigma^{-1}(\r_1)\gamma^*_\sigma(\r_1,\r_2)\gamma_\sigma(\r_1\,',\r_2\,')
\end{eqnarray}
Just as range-separated DFT requires an approximate diagonal XC
hole to determine the range-separated XC energy, projected DFT
requires an approximate nondiagonal XC hole to determine the projected XC
energy. 
 
From this perspective, Perdew \& Zunger's original orbital density
{\em{ansatz}} in eq \ref{eq:1} and the new local hybrid {\em{ansatz}} in eq
\ref{eq:2} both approximate the same quantity, the expectation value of the
nondiagonal projected electron-electron interaction in eq \ref{eq:VP} computed
with a nondiagonal approximate XC hole.  Eq \ref{eq:1} approximates this
quantity by evaluating a standard XC functional on projected one-electron
densities. Eq \ref{eq:2} approximates this quantity by locally rescaling a
standard XC functional by the ratio of projected and unprojected exact exchange
energy densities, both computed from the nondiagonal exact exchange hole.

\section{Numerical Results }

\subsection{Large-Z Limit} 

\begin{figure}
\includegraphics[width=0.7\textwidth]{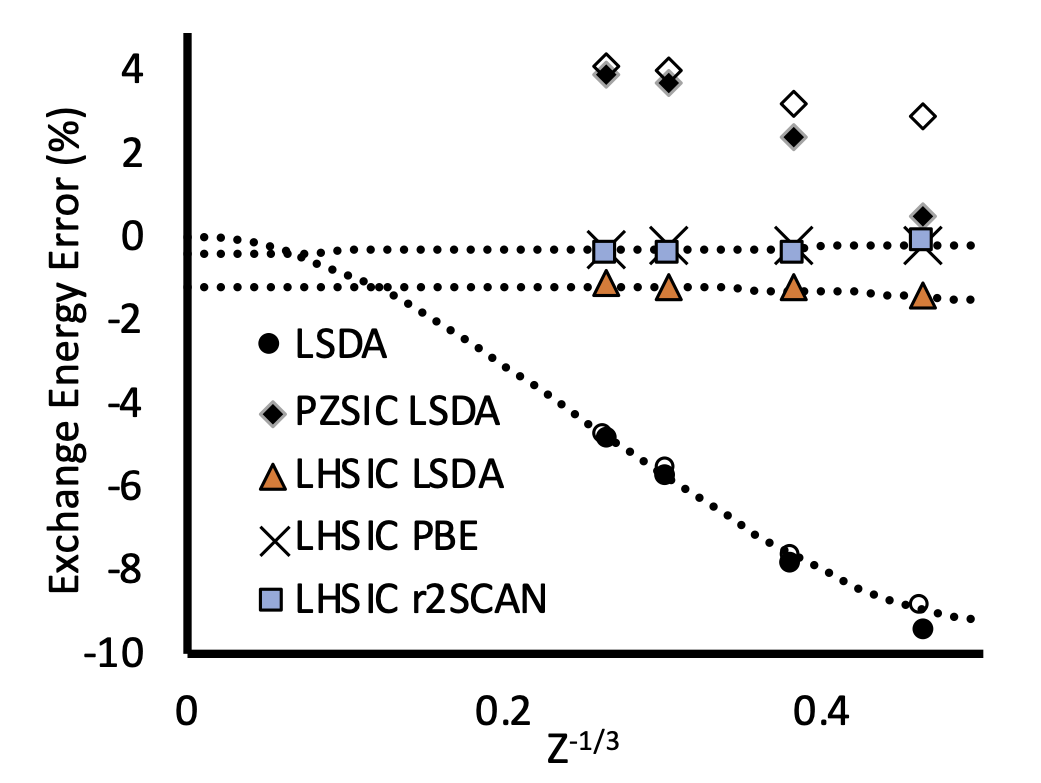}
\caption{\label{fig:ueg}Relative percent errors in neutral atom exchange energies with
nuclear charge $Z$.  Open symbols denote data from ref \cite{Santra2019a},
lines are extrapolations to the large-$Z$ limit.}
\end{figure} 

Figure \ref{fig:ueg} reports relative percent errors for exchange energies of
neutral atoms with atomic number Z=10,18,36,54, computed by applying the
Perdew-Zunger or local-hybrid-like self-interaction correction to the LSDA,
PBE, and r$^2$SCAN semilocal functionals. The figure includes least-squares
fits predicting the large-Z limit. These fits are constructed following Santra
and Perdew, who showed that these atoms' exchange energies provide a reasonable
extrapolation to the large-Z limit.\cite{Santra2019a} Reference energies are
taken from ref \cite{Chakravorty1993}. Least-square extrapolations are fit to
the functional form of ref \cite{Santra2019a} eq 13. Other computational details are in section \ref{sec:methods}. The PZ-SIC results in
Figure \ref{fig:ueg} differ slightly from the results in ref \cite{Santra2019a},
consistent with our use of Boys localized orbitals instead of fully
self-consistent FLOSIC orbitals.  The local-hybrid-like self-interaction
correction eq \ref{eq:2} gives small and nearly constant errors in
self-interaction-corrected LSDA, PBE, and r$^2$SCAN.  Ref \cite{Santra2019a}
suggested that the extrapolation gave uncertainties $\pm 0.5 \%$ in the
large-$Z$ limit, consistent with LHSIC giving zero error at large $Z$. 

\subsection{Orbital Dependence}

\begin{table}
\caption{\label{tab:l} Bond energy of Ne$_2^+$ (kcal/mol) computed with SIC-LDA
and SIC-PBE using different choices of SIC orbitals.}
\begin{tabular*}{0.75\textwidth}{@{\extracolsep{\fill}} l rr rr} 
\hline\hline 
 & \multicolumn{2}{c}{Orbital density eq \ref{eq:1}} & \multicolumn{2}{c}{Local hybrid eq \ref{eq:2} } \\ 
Orbitals & SIC-LDA & SIC-PBE & SIC-LDA & SIC-PBE \\ 
\hline
ER & 38.4 & 49.7 & 3.9 & 0.7 \\
Boys & 91.9 & 109.1 & 10.8 & -2.6 \\
KS& 233.9 & 248.8 & 5.0 & -11.7 \\
\hline\hline
\end{tabular*}
\end{table}

Table \ref{tab:l} shows the self-interation-corrected LSDA and PBE bond
energies of neon dimer cation Ne$_2^+$ at 5 Angstrom separation, computed with
various choices of SIC orbitals: Edmiston-Ruedenberg localized orbitals ER,
Boys localized orbitals, or delocalized Kohn-Sham orbitals KS. The table
compares Perdew \& Zunger's orbital density {\em{ansatz}} eq \ref{eq:1} to the
new local-hybrid-like {\em{ansatz}} eq \ref{eq:2}. The local hybrid approach
significantly reduces the orbital dependence, consistent with results reported
earliner in ref \cite{Janesko2025}. I speculate that this is a consequence
of reduced lobed density error. As discussed above, eq \ref{eq:2} does not
require evaluating the approximate XC functional on lobed one-electron
densities carved out of a many-electron density. The lobed density error thus
introduced is known to be strongly orbital-dependent.\cite{Shahi2019}

\subsection{Atomic total energies} 

\begin{table}
\caption{\label{tab:at} Mean absolute errors (MAE in Hartree) for total
energies of atoms  H-Ar.}
\begin{tabular*}{0.75\textwidth}{@{\extracolsep{\fill}} l rrr }
\hline\hline 
Method & Uncorrected & Orbital density eq \ref{eq:1} & Local hybrid eq \ref{eq:2}  \\ 
\hline 
LSDA & 0.726& 0.139 & 0.372 \\ 
PBE & 0.083 & 0.408 & 0.299 \\ 
r$^2$SCAN & 0.010 & 0.360 & 0.323 \\ 
\hline\hline
\end{tabular*}
\end{table}

Table \ref{tab:at} shows the mean absolute error in total energies of atoms
atoms H to Ar\cite{Chakravorty1993} computed with Edmiston-Reudenberg
localized orbitals. Like Table \ref{tab:l}, the table compares the orbital
density {\em{ansatz}} eq \ref{eq:1} to the local hybrid {\em{ansatz}} eq \ref{eq:2}. 
PZ-SIC with the orbital density {\em{ansatz}} eq \ref{eq:1} improves the LSDA
total energies but degrades the PBE and $r^2$SCAN total energies, consistent
with previous reports.\cite{Bhattarai2020}  The local-hybrid-like {\em{ansatz}}
eq \ref{eq:2} gives results comparable to eq \ref{eq:1} for self-interaction-corrected PBE
and r$^2$SCAN.

\subsection{Thermochemistry, Kinetics and Local scaling}

\begin{table}
\caption{\label{tab:ae} Error statistics (kcal/mol) for the W4-11 set of
thermochemical data. For example, HF-LSDA denotes exact exchange plus LSDA
correlation, LHSIC-LSDA denotes LSDA corrected by eq \ref{eq:2}.}
\begin{tabular*}{\textwidth}{@{\extracolsep{\fill}} l rr rr }
\hline\hline
Method & MD & MAD & RMSD & MaxAD \\
\hline 
LSDA            &  58.58 &   58.58 &  66.26 & 134.83 \\ 
HF-LSDA          & -53.95 &   55.22 &  67.25 & 189.04 \\ 
LHSIC-LSDA      &  -7.86 &   17.75 &  22.49 &  76.15 \\ 
\hline
PBE             &  13.24 &   14.66 &  18.09 &  51.57 \\ 
HF-PBE           & -46.40 &   46.41 &  56.18 & 165.42 \\
LHSIC-PBE       & -12.04 &   13.59 &  19.40 &  78.55 \\ 
\hline
r$^2$SCAN       &   1.09 &    3.87 &   5.26 &  24.22 \\ 
HF-r$^2$SCAN     & -48.58 &   48.59 &  58.33 & 166.01 \\ 
LHSIC-r$^2$SCAN & -19.36 &   20.35 &  26.81 &  97.24 \\ 
\hline\hline
\end{tabular*}
\end{table}

Table \ref{tab:ae} shows error statistics for the W4-11 atomization energy test set.\cite{Karton2011}
The table reports results for local-hybrid-like self-interaction-corrected LSDA, PBE and
r$^2$SCAN, as well as results for the underlying semilocal functionals and
their combinations with full exact exchange.  MD, MAD, RMSD, MaxAD denote mean
deviation, mean absolute deviation, root-mean-square deviation, and maximum
absolute deviation. 
The unscaled local-hybrid-like SIC overcorrects semilocal functionals,
exemplifying the paradox of SIC. However, the atomization energy errors are
smaller than those obtained with full exact exchange, suggesting that SIC
maintains some desirable properties of the underlying semilocal functional. In
particular, LHSIC-PBE outperforms both HF-PBE and PBE itself, motivating
further exploration. 

\begin{table}
\caption{\label{tab:llh} Error statistics (kcal/mol) for representative
datasets computed with eq \ref{eq:llh} . }
\begin{tabular*}{\textwidth}{@{\extracolsep{\fill}} l rr rr }
\hline\hline
Dataset & MD & MAD & RMSD & MaxAD \\
\hline 
     W4-11  &  -7.73 &    8.19 &   9.56 &  31.39 \\ 
     G21EA  &  -1.51 &    1.51 &   1.60 &   2.02 \\ 
     G21IP  &  -0.31 &    3.41 &   4.39 &   9.48 \\ 
      G2RC  &   0.56 &    5.12 &   6.70 &  15.71 \\ 
    BH76RC  &   0.08 &    2.63 &   4.00 &  14.87 \\ 
      BH76  &  -5.80 &    5.81 &   6.64 &  19.59 \\ 
       S22  &  -1.76 &    2.41 &   2.87 &   6.32 \\ 
    SIE4x4  &  18.45 &   18.45 &  20.59 &  37.63 \\ 
     RSE43  &  -0.87 &    1.86 &   3.23 &  16.79 \\ 
\hline\hline
\end{tabular*}
\end{table}

The paradox of self-interaction correction\cite{Perdew2015,Zope2019,Akter2021}
has been addressed by locally scaling the SIC in many-electron regions. The
local-hybrid-like formulation reported here can readily adapt scaling
strategies developed for local hybrids. Here I present a proof-of-principle
locally scalled self-interaction-corrected PBE combining a common local mixing
function and an x-LMF.\cite{Arbuznikov2024,Arbuznikov2009a}  The total energy
is 
\begin{eqnarray}
\label{eq:llh}
E &=& E_{PBE} + \int d^3\r\ a^{LLH}(\r) \left(e_{X}(\r)-e_{X}^{PBE}(\r) \right) \\ 
a^{LLH}(\r) &=& a^{xLMF}(\r) a^{SICLMF}(\r)
\end{eqnarray}
Unitless parameters $\eta=0.309$ and $\vartheta=3.209$ for the x-LMF are taken
from ref \cite{Arbuznikov2024}. The unmodified PBE exchange energy density is
used in the x-LMF eq \ref{eq:xlmf}. All exchange energy densities energies are
spin-summed. 
 
Table \ref{tab:llh} reports error statistics obtained with this approach for
representative test sets of molecular thermochemistry, kinetics, and relevant
properties. Test sets are taken from the GMTKN55 dataset.\cite{Goerigk2017} The
selected test sets include W4-11 atomization energies, G21IP and G21EA
$\Delta$SCF ionization potentials and electron affinities, G2RC and BH76RC
reaction energies, BH76 reaction barriers, S22 noncovalent interaction
energies, and the SIE4x4 self-interaction-error problems and RSE43 reaction
energies.  ionization potentials, PA26 proton affinities, SIE4x4
self-interaction-error problems and RSE43 radical stabilization energies. 
The MAE in total energies of the atoms H-Ar is 0.049 Hartree, below that for
PBE in table \ref{tab:at}. 
Eq \ref{eq:llh} improves upon uncorrected PBE for most of the t
properties in Table \ref{tab:llh}. Eq \ref{eq:llh} gives particularly good results for the RSE43 set
of radical stabilization energies and the SIE-4x4 set of radical cation bond energies,
test cases where self-interaction error is known to play a significant
role.
Excessively repulsive noncovalent interactions are a ``smoking gun'' for energy
gauge errors in local hybrid functionals. The x-LMF in eq \ref{eq:llh}
mitigates such errors in local hybrids.\cite{Arbuznikov2024} Table \ref{tab:llh}
suggests that the x-LMF also helps mitigate such errors in local-hybrid-like
SIC. For the S22 set, eq \ref{eq:llh} gives MAD 2.41 kcal/mol, somewhat larger
than PBE (MAD 1.89 kcal/mol) and well below SIC-PBE (MAD 3.25 kcal/mol),
suggesting that energy gauge errors are at least mitigated. 
Eq \ref{eq:llh} gives W4-11 atomization energy errors that are still well above
r$^2$SCAN, unacceptably high for a modern density functional. 
The local scaling in eq \ref{eq:llh} does not transfer to
self-interaction-corrected r$^2$SCAN. Replacing PBE with r$^2$SCAN in eq
\ref{eq:llh} gives an S22 MAD above 4 kcal/mol and a W4-11 MAD above 30
kcal/mol. These facts motivate further development.

\section{Discussion} 

Self-interaction error and its correction are foundational concepts in density
functional theory. Practical applications of the Perdew-Zunger self-interaction
correction are limited by its errors for uniform systems.\cite{Santra2019a} The
local-hybrid-like self-interaction-correction {\em{ansatz}} LHSIC discussed
here restores the uniform limit and mitigates lobed density errors, opening new
possibilities for practical applications of SIC. The LHSIC clearly
demonstrates how self-interaction corrections lie on the fourth rung of Jacob's
Ladder of density functional approximations\cite{Perdew2001} alongside global,
local, and range-separated hybrid functionals and other
hyper-GGAs.\cite{Perdew2008,Haunschild2012,Shahi2019}

The results presented here suggest multiple possibilities for future work. 

{\em{Efficient Implementation}}. Modern local hybrid codes such as
TURBOMOLE\cite{Franzke2023} offer efficient local hybrid calculations of a wide
range of properties, from magnetic response to excited
states.\cite{Maier2018,Grotjahn2020a,Grotjahn2019,Fuerst2023a} The local hybrid
formulation in eq \ref{eq:siclmf} should be readily integrated into these
codes. The reduced orbital dependence dependent in Table \ref{tab:l} means that
such implementations could at least initially use Foster-Boys,
Edmiston-Reudenberg, or other efficient orbital localization
procedures\cite{Peralta2026} in lieu of the more expensive FLOSIC procedure. 

{\em{Constraint Satisfaction}}. One goal of self-interaction correction is a
``do no harm'' approach\cite{Shahi2026} that removes SIE without degrading the
broad accuracy of nonempirical meta-GGAs.  The local hybrid formulation of SIC
opens new possibilites for achieving this dream.  LHSIC should be readily
applied to the nonempirical meta-GGA local hybrids developed by Holzer and
Francke.\cite{Holzer2022,Holzer2024} These authors use the Tao-Mo density
matrix expansion\cite{Tao2016} to combine constraint satisfaction with a
minimal gauge mismatch between exact and semilocal exchange. Simply replacing
the correlation lengh LMF\cite{Johnson2014} in CYHF-PBE\cite{Holzer2024} with
eq \ref{eq:siclmf} should provide an interesting test of constraint
satisfaction within SIC. An alteranative is to revisit self-interaction-corrected
r$^2$SCAN combining eq \ref{eq:siclmf} with a partial integration gauge
correction.\cite{Maier2016a} LHSIC should also enable construction of more
flexible local scaling approaches, as it does not require the local scaling to
got to zero in uniform systems.\cite{Shahi2026} 

{\em{Nondiagonal Approximate XC Holes}}. When the SIC is derived as in section
\ref{proj}, the Perdew-Zunger orbital density approach and the
local-hybrid-like approach both become approximations for a projected XC energy
obtained from a nondiagonal projected XC hole.\cite{Janesko2025} This implies
that further progress should be possible by explicitly modeling the nondiagonal
XC hole. Such models may draw on earlier work in range-separated DFT, in which
diagonal model exchange holes were constructed to reproduce existing exchange
functionals.\cite{Iikura2001,Bahmann2008,Henderson2008,Ernzerhof1998} One
promising direction is to construct nondiagonal exchange holes from the Tao-Mo
functional, inserting a density matrix expansion into eq
\ref{eq:nondiag}.\cite{Tao2016} Precedents for this work include rung-3.5
density functionals constructed from the density matrix expansion (and thus
implicitly from a nondiagonal approximate exchange
hole)\cite{Ramos2020,Ramos2021} as well as nondiagonal correlation holes
obtained from relatively low-cost regularized perturbation
theory.\cite{Janesko2026} It remains an open question whether explicit
nondiagonal model exchange holes could mitigate the energy gauge effects seen
in local-hybrid-like SIC. 

{\em{Self-And-Some-Others Correction: Projected Configuration Interaction
Density Functional Theory}}. In my view, the most promising future direction is
to introduce low-cost wavefunction correlation. We've extended the derivation
in section \ref{proj} to a ``self-and-some-others'' correction, simply by
projecting the electron-electron interaction onto additional SIC
orbitals.\cite{Janesko2022b} To illustrate the potential of this approach,
consider a closed-shell system possessing $N/2$ doubly occupied localized SIC
orbitals and $N_v=N_{AO}-N/2$ unoccupied (virtual) orbitals. One could
introduce a ``pair interaction correction'' built from $N/2$ different
projected-interacting reference systems. The $i$th reference system would
introduce self-interaction {\em{and}} pair correlation for the $i$th electron
pair, by projecting the electron-electron interaction onto the $i$th localized
occupied orbital and all $N_v$ delocalized virtual orbitals.  Numerically exact
correlation energies for each projected-interacting reference systems could be
obtained from $N/2$ different two-electron $(N_V+1)$-orbital CI calculations.
Correcting these reference system correlation energies with exact projected XC
functionals, constructed from the nondiagonal XC holes discussed above, would
ensure that any weighted sum over reference system energies would yield the
exact ground state energy without the need to localize the virtual
orbitals.\cite{Janesko2025} (Specialists may note that the reference systems
all use the same virtual orbitals, such that the sum of reference system
correlation energies could be larger in magnitude than the exact correlation
energy. In such systems, the exact projected correlation functionals would
yield a positive correction.)  Work in this direction is underway.

\section{Computational Details} 
\label{sec:methods} 

All calculations use a standalone implementation of local-hybrid-like
self-interaction correction in PySCF.\cite{Sun2020} The code is freely
available online at github.com/bjanesko/ProjectedInteractingDFT. 
All energies are computed non-self-consistently using Hartree-Fock or
B3LYP\cite{Cheeseman1996,Becke1993,Becke1988,Lee1988} orbitals and orbital
energies.
The exact exchange energy density is computed using a resolution of the
identity in the atomic orbital basis set:\cite{Jaramillo2003}
\begin{eqnarray}
e_{X\sigma}(\r) &=& \frac{1}{4}\sum_{\mu\nu\eta\lambda}\chi_\mu(\r)S^{-1}_{\mu\nu}K_{\sigma,\nu\eta}P_{\sigma,\eta\lambda}\chi_\lambda(\r) \textrm{+ c.c.} \\ 
K_{\sigma,\mu\nu}&=&-\sum_{\lambda\eta}\left<\mu\lambda|\nu\eta\right>P_{\sigma,\lambda\eta}
\end{eqnarray}
Calculations on the heavy atoms in Figure \ref{fig:ueg}  and test calculations
in Table \ref{tab:l} use Foster-Boys localized occupied orbitals for
self-interaction correction.\cite{Boys1970} Other calculations use
Edmiston-Ruedenberg localized orbitals for SIC.\cite{Ruedenberg1982}
Calculations in Figure \ref{fig:ueg} use Dyall's valence-quintuple-zeta basis
sets as distributed on the Basis Set
Exchange.\cite{dyall2026a,schuchardt2007a,feller1996a,pritchard2019a} 
Calculations in Figure \ref{fig:ueg} and Tables \ref{tab:l}-\ref{tab:at} are
post-Hartree-Fock. Calculations in Tables \ref{tab:ae}-\ref{tab:llh} use B3LYP
orbitals and orbital energies computed in the def2QZVP basis
set\cite{Weigend2005} using Gaussian 16.\cite{g16}
The latter calculations are intentionally not performed post-Hartree-Fock to
avoid ``too good'' barriers from reduced density-driven
error.\cite{Wasserman2017} 
Datasets and reference values are taken from the GMTKN55
database.\cite{Goerigk2017} 

%

%
%

\printbibliography

\end{document}